\documentclass[11pt]{llncs}

\usepackage{latexsym}
\usepackage{graphicx}
\usepackage{ascmac}
\usepackage{algorithm,algorithmic}
\usepackage{color}
\usepackage{url}

\usepackage{amsmath, amssymb}
\usepackage{moreverb}
\usepackage{comment}

\graphicspath{{.}{./figure/}}

\newcommand{\MLCMPP}{MLCM-P\_PATH }
\newcommand{\CIC}{CIRCLE INSIDE CROSSING }

\renewcommand{\baselinestretch}{1.00}

\makeatletter
\spn@wtheorem{fact}{Fact}{\bfseries}{\itshape}
\makeatother

\makeatletter
\renewcommand{\ALG@name}{}
\renewcommand{\listalgorithmname}{List of \ALG@name s}
\makeatother

\newcommand{\nop}[1]{}

 \makeatletter
 \long\def\@caption#1[#2]#3{\par\addcontentsline{\csname
   ext@#1\endcsname}{#1}{\protect\numberline{\csname 
   the#1\endcsname}{\ignorespaces #2}}\begingroup
     \@parboxrestore
     \small
     \@makecaption{\csname fnum@#1\endcsname}{\ignorespaces #3}\par
   \endgroup}
 \makeatother

\title{Exact and fixed-parameter algorithms for metro-line crossing minimization problems} 

\author{%
   Yoshio Okamoto\inst{1} \and
   Yuichi Tatsu\inst{2} \and
   Yushi {\rm Uno}\inst{2}
}

\institute{
    Department of Communication Engineering and Informatics, 
    Graduate School of Informatics and Engineering, 
    The University of Electro-Communications, 
    1-5-1 Chofugaoka, Chofu, Tokyo 182-8585, Japan. 
    \email{okamotoy@uec.ac.jp} 
\and
    Department of Mathematics and Information Sciences, 
    Graduate School of Science, Osaka Prefecture University, 
    1-1 Gakuen-cho, Naka-ku, Sakai 599-8531, Japan. 
    \email{sr301023@edu.osakafu-u.ac.jp,uno@mi.s.osakafu-u.ac.jp}
}

\begin{document}

\maketitle

\begin{abstract}
A metro-line crossing minimization problem 
is to draw multiple lines on an underlying graph that models stations 
and rail tracks so that the number of crossings of lines becomes minimum. 
It has several variations by adding restrictions on how lines are drawn. 
Among those, there is one with a restriction that line terminals 
have to be drawn at a verge of a station, 
and it is known to be NP-hard even when underlying graphs are paths. 
This paper studies the problem in this setting, 
and propose new exact algorithms. 
We first show that a problem to decide if lines can be drawn without crossings 
is solved in polynomial time, and propose a fast exponential algorithm 
to solve a crossing minimization problem. 
We then propose a fixed-parameter algorithm 
with respect to the multiplicity of lines, 
which implies that the problem is FPT\@. 
\end{abstract}

\section{Introduction}

A visual representation of graphs
greatly helps to understand 
what they model and stand for, 
and one of the objectives of graph drawing research is 
to design such algorithms that generate 
informative drawings. 
Among several targets to draw, metro maps
have attracted much research interest, 
and the 
quality of drawn maps are often measured 
by the number of crossings of metro-lines and the goal is to minimize it. 
To model this, a graph is defined by regarding stations and rail tracks 
as vertices and edges, respectively, 
and by regarding a metro-line as a path on the graph, 
those paths are required to be drawn with fewer crossings. 
This model consists of two problems; 
one to draw a graph with fewest crossings of edges on the plane, 
and the other to draw paths with fewest crossings 
on a fixed drawing of a graph. 
The former is a classic problem well known as crossing number minimization of graphs, 
but the latter is a recently proposed problem 
as metro-line crossing minimization problem (MLCM) 
by Benkert et al. \cite{benkert07}. 

This paper studies a variation of MLCM 
with a restriction that line terminals 
have to be drawn at a verge of a station, which is called MLCM-P\@. 
It is known to be NP-hard even when its underlying graphs are paths, 
and we focus on such cases of MLCM-P\@. 
We first show that a problem to decide if lines can be drawn 
without crossings can be solved in polynomial time
by reducing it to the graph planarity problem. 
Then we propose a fast exponential-time exact algorithm 
to solve a crossing minimization problem 
by utilizing the properties of the possible relative positions of lines. 
N\"{o}llenburg \cite{nollenburg10} posed as an open problem for MLCM-P 
if it is fixed-parameter tractable,
and also pointed out that the multiplicity of lines is a possible parameter. 
From this point of view, 
we propose a fixed-parameter algorithm with respect to 
this parameter based on dynamic programming approach.
This is designed by carefully observing properties 
of optimal solutions of MLCM-P\@. 
This result partially solves his open problem affirmatively, 
and such an algorithm should run practically fast 
since the multiplicity of lines is considered to be small in reality. 

In Sect.~\ref{MLCM} we give terminology and definitions for MLCM, 
and explain MLCM-P 
with its related research results. 
Sect.~\ref{MLCM-P_crossing} shows that MLCM-P, 
which is NP-hard in general, is polynomially solvable 
for deciding if lines are drawn without crossings 
when underlying graphs are restricted to paths. 
Sect.~\ref{MLCM-P_exact} gives a fast exponential algorithm 
for MLCM-P and Sect.~\ref{MLCM-P_fpt} establishes a fixed-parameter algorithm 
by introducing multiplicity as its parameter, 
both for MLCM-P when underlying graphs are paths. 

\section{Metro-line Crossing Minimization Problems}
\label{MLCM}

This section gives terminology for modeling metro-line problems 
and an overview of related research results. 

\subsection{MLCM and the Periphery Condition}

\renewcommand{\thefootnote}{\fnsymbol{footnote}}
\setcounter{footnote}{3}

\begin{figure}[htbp]
 \begin{center}
\scalebox{0.65}{\includegraphics{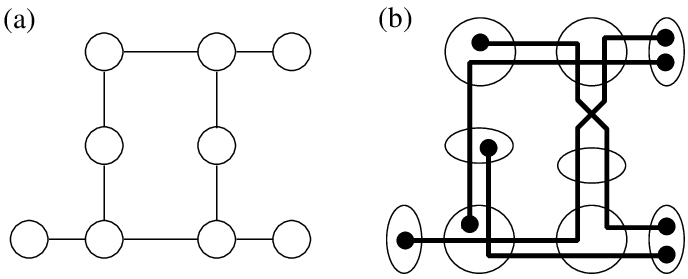}}
 \end{center}
\caption{(a) An underlying graph, and 
(b) lines drawn on the underlying graph whose edges are omitted.} 
\label{underlying_graph}
\end{figure}

An \emph{underlying graph} of a metro map is a graph $G=(V,E)$ 
embedded on the plane 
whose vertex set $V$ and edge set $E$ denote stations 
and rail tracks between two stations, respectively. 
In metro-line crossing minimization problems, 
we assume that a drawing of an underlying graph is fixed as an input 
(Fig.~\ref{underlying_graph}(a)). 
Each metro-line to be drawn on the underlying graph $G$ 
is called a \emph{line}, and is assumed to be a simple path on $G$. 
For a line $l_i=(v_{i_0},v_{i_1},\ldots, v_{i_k})$, 
vertices $v_{i_0}$ and $v_{i_k}$ are its {\it ends}. 
We denote by ${\cal L}$ the set of lines to be drawn, 
and we assume that lines $l_i$ in ${\cal L}$ are distinct. 
In drawing metro maps in reality, we assume that each station has its area, 
like a rectangle or an oval for example, 
and we usually omit drawing the edges in $E$ 
(Fig.~\ref{underlying_graph}(b)). 

For edge $(u,v)\in E$, we denote a set of lines in ${\cal L}$ 
that pass along $(u,v)$ by ${\cal L}_{uv}$ ($={\cal L}_{vu}$, by definition). 
We define the {\it order} $<^u_{uv}$ $(<^v_{uv})$ 
of two lines $l_1,l_2\in{\cal L}_{uv}$ at each end of edge $(u,v)$ 
by regarding $(u,v)$ as a directed arc, that is, 
when we see $v$ from $u$ 
we say that $l_1<^u_{uv}l_2$ ($l_1<^v_{uv}l_2$) 
if $l_1$ lies on the right side of $l_2$ on $u$ ($v$). 
Then two lines $l_1$ and $l_2$ cross on edge $(u,v)$ 
if $l_1<^u_{uv}l_2$ and $l_2<^v_{uv}l_1$ 
(or $l_2<^u_{uv}l_1$ and $l_1<^v_{uv}l_2$) hold 
(see Fig.~\ref{uv_lines} in Appendix). 
We denote by $s^u_{uv}$ ($s^v_{uv}$) the sequence of lines in a line subset 
${\cal L}_{uv}$ sorted according to the order $<^u_{uv}$ ($<^v_{uv}$). 
To draw a line subset $A\subseteq{\cal L}$ on an underlying graph $G$ 
is to determine $s^u_{uv}$ and $s^v_{uv}$ with respect to the lines in $A$ 
on every edge $(u,v)\in E$, and it defines a {\it layout} of $A$. 
A station is supposed to be drawn with painting its inside, 
and therefore, we make an assumption 
that lines which come into vertex $v$ from $u$ have to go out 
along edges incident to $v$ (except $(u,v)$) without crossing inside of $v$. 
Formally, let edges incident to $v$ (except $(u,v)$) 
be $vw_1,\ldots,vw_k$ in counterclockwise order. 
Then the sequence formed by deleting the lines one of whose end is $v$ 
from $s^v_{uv}$ must be a subsequence of the sequence 
$s^v_{vw_1},\ldots$, $s^v_{vw_k}$ concatenated in this order. 
When this condition holds for all edges incident to $v$, 
such a vertex is called {\it admissible} 
(see Fig.~\ref{admissible} in Appendix). 
If all the vertices of a graph are admissible, 
such a layout is also called admissible. 
For an admissible layout of a line set ${\cal L}$ on an underlying graph $G$, 
its {\it crossing number} is the total number of crossings of lines 
that occur on the edges of the graph. 
%

\begin{figure}[htbp]
 \begin{center}
\scalebox{0.65}{\input{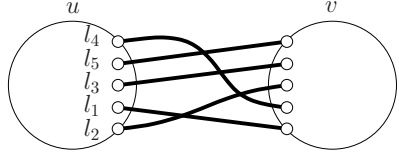}}
 \end{center}
 \caption{Lines on edge $(u,v)$.}
 \label{uv_lines}
\end{figure}

\vspace*{-0.5cm}
\begin{figure}[htbp]
 \begin{center}
\scalebox{0.65}{\includegraphics{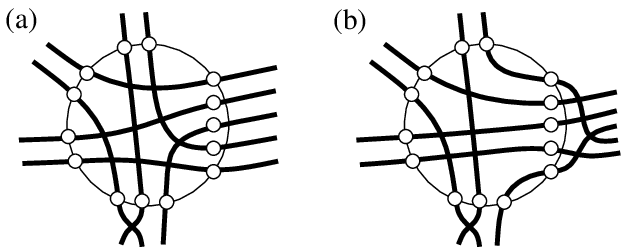}}
 \end{center}
 \caption{Two vertices (stations) 
that are (a) not admissible, and (b) admissible.} 
 \label{admissible}
\end{figure}

Now the problem MLCM is formally defined as follows \cite{benkert07}. 
\begin{quote}
{\bf MLCM}

Input: an underlying graph $G$ (embedded on the plane) 
and a line set ${\cal L}$ on $G$. 

Output: an admissible layout of ${\cal L}$ whose number of crossings 
is minimum. 
\end{quote}

\begin{figure}[htbp]
 \begin{center}
\scalebox{0.60}{\includegraphics{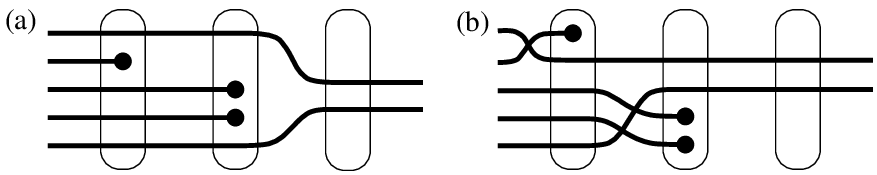}}
 \end{center}
\caption{Two layouts: (a) one that does not satisfy, 
and (b) the other that satisfies the periphery condition.} 
 \label{periphery}
\end{figure}

\noindent
In an output of MLCM, it is allowed that an end of a line in a station 
is drawn between lines that pass through that station. 
However, recent research proposes a model that forbids such drawings 
since they may introduce some confusion \cite{nollenburg10}. 
Such a restriction (on a line end) is called the {\it periphery condition} 
and is formally defined as follows: 
an end $v_{i_0}$ ($v_{i_k}$) of line $l_i=(v_{i_0},\ldots, v_{i_k})$ 
satisfies the periphery condition 
if $l_i$ appears either in a leftmost or rightmost position with respect to 
$<^{v_{i_0}}_{v_{i_0}v_{i_1}}$ ($<^{v_{i_k}}_{v_{i_{k-1}}v_{i_k}}$) 
among all the lines in 
${\cal L}_{v_{i_0}v_{i_1}}$ (${\cal L}_{v_{i_{k-1}}v_{i_k}}$) 
except the ones whose end is $v_{i_0}$ ($v_{i_k}$) (Fig.~\ref{periphery}). 
A variation of MLCM 
where each line of its output satisfies the periphery condition 
is called MLCM-P (P stands for periphery), and is described as follows \cite{bekos08}. 

\begin{quote}
{\bf MLCM-P}

Input: an underlying graph $G$ and a line set ${\cal L}$ on $G$. 

Output: an admissible layout of ${\cal L}$ where each line satisfies 
the periphery condition, and whose crossing number is minimum. 
\end{quote}

\noindent
When we determine the position of line $l=(v_{i_0},v_{i_1},\ldots,v_{i_k})$ 
to be leftmost (rightmost) with respect to $<^{v_{i_0}}_{v_{i_0}v_{i_1}}$ 
in ${\cal L}_{v_{i_0}v_{i_1}}$ (so that its end $v_{i_0}$ satisfies 
the periphery condition), 
we say that we {\it assign} the end $v_{i_0}$ to left (right) 
(similarly to $v_{i_k}$). 
On the other hand, a problem to find a layout 
under the condition that such assignments are given as inputs 
is proposed as MLCM-PA (A stands for assignment) \cite{bekos08}. 
\begin{quote}
{\bf MLCM-PA}

Input: an underlying graph $G$, a line set ${\cal L}$ on $G$ 
and assignments (to right or left) of both ends of the lines in ${\cal L}$. 

Output: an admissible layout of ${\cal L}$ whose crossing number is minimum, 
where the position of each line end satisfies the given assignment. 
\end{quote}

\subsection{Related Results and Our Problem Setting}

This subsection introduces the related results 
on MLCM and its variations MLCM-P and MLCM-PA. 
Benkert, N\"{o}llenburg, Uno and Wolff \cite{benkert07} 
started to study the problem MLCM, 
and proposed an $O(|{\cal L}|^2)$ algorithm that determines an order of lines 
for an edge when underlying graphs are planar. 
Bekos, Kaufmann, Potika and Symvonis \cite{bekos08} showed 
that MLCM-P is NP-hard even when underlying graphs are paths, 
and proposed an $O(|{\cal L}||V|)$ algorithm for MLCM-PA 
when underlying graphs are paths or trees and every station 
are restricted to have at most two sides (2-side model). 
Asquith, Gudmundsson and Merrick \cite{asquith08} 
designed an $O(|{\cal L}|^3|E|^{2.5})$ algorithm for MLCM-PA 
when underlying graphs are planar. 
For MLCM-P, they proposed a method to obtain an optimal assignment 
of each line end first by formulating and solving it 
as an integer program and then solving MLCM-PA by giving 
those assignments as inputs. 
Argytiou, Bekos, Kaufmann and Symvonis \cite{argytiou09} proposed 
an $O(|V|(|E|+|{\cal L}|))$ algorithm for MLCM-PA when underlying graphs 
are planar and stations are restricted to 2-side model, and 
N\"{o}llenburg \cite{nollenburg10} showed an $O(|{\cal L}|^2|V|)$ algorithm 
for MLCM-PA when underlying graphs are planar. 
Moreover, N\"{o}llenburg \cite{nollenburg10} listed some open problems, e.g., 
that ask if MLCM is NP-hard for planar graphs, 
if MLCM-P is fixed-parameter tractable
and if approximation algorithms for MLCM-P exist. 
As we saw, for metro-line crossing minimization problems, 
the main concern so far was to design efficient algorithms for MLCM-PA 
which is tractable, 
and not so many results can be seen for MLCM and MLCM-P\@. 

Now, in this paper, we discuss MLCM-P which is intractable in general, 
and focus on the case that its underlying graphs are paths, 
which is still intractable. 
We name this problem setting \MLCMPP for short. 
In the subsequent discussions, 
an input underlying graph $G$ is always a path, and therefore 
we express its vertex set by $V(G)=\{1,2,\ldots, n\}$ and edge set 
by $E(G)=\{(i, i+1)\mid 1\leq i\leq n-1\}$, 
and a line is denoted by a pair of its two ends as $l=[i, j]$ ($i<j$) 
instead of a sequence of vertices. 
We call $i$ and $j$ left and right end of line $l$, respectively. 
For simplicity, we assume that an underlying path is a horizontal line segment 
and that its left (right) end corresponds to station 1 ($n$). 
Therefore, stations become two-sided, 
and for each station $i$ we call the side to which station $i-1$ ($i+1$) 
exists the left (right) side. 
Also for intuition, when we see station $i+1$ from $i$, 
we say that we assign the end $i$ ($j$) of line $l=[i,j]$ to top (or bottom) 
instead of left or right, respectively.

\section{\MLCMPP Crossing Problem}
\label{MLCM-P_crossing}

Even when a minimization problem is NP-hard, 
there are some cases where its decision version, 
e.g., to determine whether its minimum value is 0, belongs to class P\@. 
For example, a problem to compute the minimum number of crossings $cr(G)$ 
when a graph $G$ can be drawn on the plane 
(CROSSING NUMBER) is NP-hard, while a problem to ask if $cr(G)=0$ (PLANARITY) 
is solvable in linear time \cite{hopcroft+tarjan74}. 
Also a problem to draw binary tanglegrams with minimum number of crossings 
is NP-hard, while to ask if it is 0 is solved in linear time 
\cite{fernau05}. 
Although MLCM-P\_PATH is NP-hard, 
we can consider its decision version, that is, 
to determine if the number of crossings of its optimal layout is 0, 
which we name \MLCMPP CROSSING, 
and we have the following fact for this problem. 

\begin{theorem}
\label{MLCM-PP_crossing_in_P}
{\rm
\MLCMPP CROSSING is in P\@. 
}
\end{theorem}

We will prove Theorem~\ref{MLCM-PP_crossing_in_P} 
by reducing \MLCMPP CROSSING to PLANARITY. 
To this end, we introduce the following artificial problem, 
{\sc CIRCLE INSIDE CROSSING} (CIC), and take two steps: 
first reduce \MLCMPP CROSSING to CIC, 
and then reduce CIC to PLANARITY. 

\begin{quote}
{\bf CIRCLE INSIDE CROSSING (CIC)}

Input: a graph $H=(V,E)$ and a bijection $\delta\colon V\longrightarrow 
\{1,2,\ldots,|V|\}$. 

Output: draw vertices of $V$ on a single line in the order defined by $\delta$ 
and a circle that passes through $\delta^{-1}(1)$ and contains 
all the other vertices: 
then if all the edges in $E$ are drawn within the circle 
without crossings, output yes; otherwise no. 
\end{quote}

\begin{figure}[htbp]
 \begin{center}
\scalebox{0.65}{\input{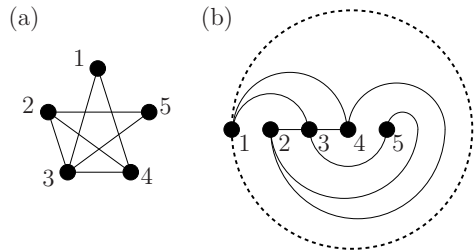}}
 \end{center}
\caption{(a) An input graph for CIC, 
and (b) its drawing within a circle without crossings.} 
 \label{circle_inside}
\end{figure}

\noindent
For example, for a graph shown in Fig.~\ref{circle_inside}(a) 
and $\delta(i)=i$, since there exists a drawing 
shown in Fig.~\ref{circle_inside}(b), the output is yes. 

\medskip
\noindent
{\bf First Step: }
In the first step, we transform an instance $I=(G,{\cal L})$ 
of \MLCMPP CROSSING 
to an instance $I'=(G',\delta')$ of CIC in the following manner. 
Remember that $G$ in $I$ is defined by $V(G)=\{1,\ldots, n\}$ 
and $E(G)=\{(i,i+1)\mid 1\leq i < n\}$, and $l\in {\cal L}$ is defined 
by $[i,j]$.
We define $G'$ by letting $V(G')=V(G)$ 
and $E(G')=E(G)\cup\{(i,j)\mid [i,j]\in{\cal L}\}$. 
We also define $\delta'(i)=i$. 
For an instance $I'$ obtained from $I$ in this way, 
we have the following lemma. 

\begin{lemma}
\label{I=I'}
{\rm
The minimum number of crossings of an instance $I$ of \MLCMPP CROSSING 
and of an instance $I'$ of CIC are equal. 

\begin{proof}
Once we determine an admissible layout for an instance $I$ of MLCM-P, 
we have its corresponding assignments to top or bottom 
of both ends of lines in ${\cal L}$. 
Let such an assignment be ${\cal A}$ and let an instance of MLCM-PA 
be $(G,{\cal L},{\cal A})$ 
defined by $I=(G,{\cal L})$ together with ${\cal A}$. 
Below, to prove Lemma~\ref{I=I'}, 
we construct a graph $G^*=(V^*, E^*)$ from an instance 
$(G,{\cal L}, {\cal A})$ of MLCM-PA and observe its properties. 
Let $V^*=\{1_\uparrow, 1_\downarrow,\ldots, n_\uparrow, n_\downarrow \}$, 
where $i_\uparrow$ and $i_\downarrow$ correspond to top and bottom, 
respectively, of each vertex $i$ of $V(G)$, 
and let $E^*$ be an edge set each of whose edge connects two vertices of $V^*$ 
corresponding to both ends of each line in ${\cal L}$ 
with assignment to top or bottom of its ends. 
For $G^*$ constructed in this way, a {\it circular drawing} of $G^*$ is 
to draw a circle and put vertices $1_\uparrow, 2_\uparrow, \ldots, 
n_\uparrow, n_\downarrow, (n-1)_\downarrow, \ldots, 1_\downarrow$ 
in this order on the circle counterclockwise, 
and to draw each edge in $E^*$ as a chord to connect two end vertices 
on a circle (Fig.~\ref{I->I'}(a) and (b)). 

\begin{figure}[htbp]
 \begin{center}
\scalebox{0.62}{\input{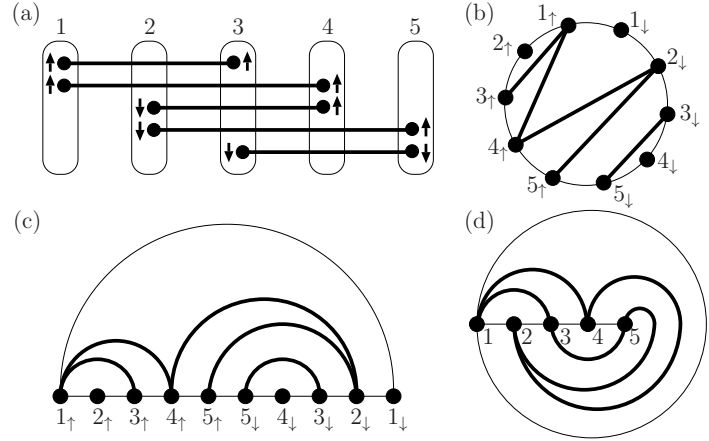}}
 \end{center}
 \caption{
(a) A layout of $I$ of MLCM-P without crossings and its corresponding instance 
$(G,{\cal L}, {\cal A})$ of MLCM-PA, 
(b) a circular drawing of $G^*$ without crossings constructed 
from $(G,{\cal L},{\cal A})$, 
(c) a transformation into a half-circle drawing of $I'$, 
(d) a circle-inside drawing of $I'$.} 
\label{I->I'}
\end{figure}

Then the following lemma holds. 

\begin{lemma}
\label{circular_drawing}
{\rm
The minimum number of crossings of an instance $(G,{\cal L}, {\cal A})$ 
of MLCM-PA and the crossing number of the circular drawing of $G^*$ are equal. 

\begin{proof}
By transforming a layout with minimum number of crossings of an instance 
$(G,{\cal L}, {\cal A})$ of MLCM-PA in the following manner, 
we will obtain a circular drawing of $G^*$ with the same number 
of crossings. 
For such a layout of MLCM-PA (Fig.~\ref{circular_drawing_transform}(a)), 
we first generate vertices $i_\uparrow$ and $i_\downarrow$ 
that correspond to top and bottom of each station $i$, 
and locate them to the top and the bottom of station $i$, respectively. 
Then connect the ends of lines that are assigned to top or bottom 
at either side of station $i$ to $i_\uparrow$ or $i_\downarrow$, respectively, 
with keeping their relative positions. 
Further, draw a rectangle to make its perimeter pass vertices 
$1_\uparrow,\ldots$, $n_\uparrow$, $1_\downarrow,\ldots$, $n_\downarrow$ 
in this order and to contain all stations and lines in it 
(Fig.~\ref{circular_drawing_transform}(b)). 
Next remove ovals that represent stations, and transform the circumference 
of the rectangle continuously into a circle by keeping relative positions 
and connections of vertices, lines, and their crossings 
(Fig.~\ref{circular_drawing_transform}(c)). 

Now in addition to vertices 
$1_\uparrow,\ldots$, $n_\uparrow$, $1_\downarrow,\ldots$, $n_\downarrow$, 
by viewing line ends as vertices and connectors between vertices as edges, 
we regard all of these elements to constitute a graph. 
In this graph, contract vertex $i_\uparrow$ ($i_\downarrow$) 
and those connected to it into one, and then draw all the edges 
as a straight line segment (chord of a circle). 
Finally, flip the entire graph by axis line $1_\uparrow n_\downarrow$, 
and we obtain a circular drawing of $G^*$ whose number of crossings 
is equal to the minimum number of crossings of a layout of MLCM-PA 
(Fig.~\ref{circular_drawing_transform}(d)). 

Conversely, by doing this transformation in the reverse order, 
we obtain a layout of MLCM-PA whose number of crossings is the same 
as that in a circular drawing of $G^*$. 
\qed

\begin{figure}[htbp]
 \begin{center}
\scalebox{0.62}{\input{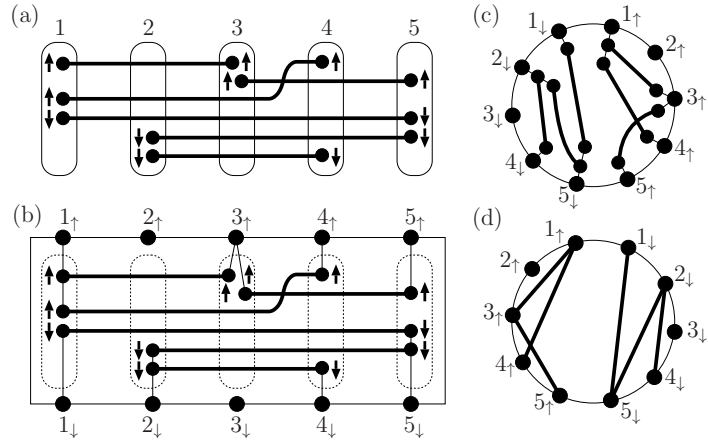}}
 \end{center}
 \caption{(a) A layout of an instance for MLCM-PA 
with minimum number of crossings, 
(b) draw a surrounding rectangle, add vertices on it and connectors to them, 
(c) remove stations and transform the rectangle into a circle, and 
(d) contract each set of vertices related to $i_\uparrow$ ($i_\downarrow$).} 
\label{circular_drawing_transform}
\end{figure}
\end{proof}
}
\end{lemma}

Now we can see that the
assignment of each end of lines is determined 
once we fix a layout with minimum number of crossings 
for an instance $I$ of \MLCMPP CROSSING. 
So let $(G,{\cal L}, {\cal A}^*)$ be an instance of MLCM-PA, 
whose input is defined by $I=(G,{\cal L})$ 
together with such assignment ${\cal A}^*$. 
By Lemma~\ref{circular_drawing}, 
the number of crossings in a circular drawing of $G^*$ constructed from 
$(G,{\cal L}, {\cal A}^*)$ coincides with the minimum number of crossings 
of $(G,{\cal L}, {\cal A}^*)$. 
Then we obtain a drawing with the same number of crossings 
of an instance $I'$ of CIC 
by transforming a circular drawing of $G^*$ in the following manner. 
For a circular drawing of $G^*$, put its vertices 
$1_\uparrow$, $2_\uparrow,\ldots$, $n_\uparrow$, $n_\downarrow$, 
$(n-1)_\downarrow,\ldots$, $1_\downarrow$ on a straight line 
from left to right in this order, and `extend' each chord and a circle 
in the form of half-circle accompanied by this operation 
(Fig.~\ref{I->I'}(c)). 
Setting the midpoint of $n_\uparrow$ and $n_\downarrow$ to be a center, 
we `fold' the part of $n_\downarrow,\ldots$, $1_\downarrow$ 
by rotating it $180^\circ$ clockwise, 
and contract each $i_\uparrow$ and $i_\downarrow$ to make vertex $i$ 
(Fig.~\ref{I->I'}(d)). 
Such a drawing (of $I'$) obtained in this way 
has the same number of crossings as the minimum number of crossings 
of $(G,{\cal L}, {\cal A}^*)$, 
by regarding the arc segments of a drawing of $G^*$ 
as parts of $E(G')$ of $I'$ of CIC. 

Conversely, starting from a drawing with minimum number of crossings of $I'$, 
we have a circular drawing of $G^*$ with the same number of crossings 
by executing this transformation in the reverse direction. 
Then we have a layout of $I$ with the same number of crossings 
by making assignments of ends of lines in ${\cal L}$ of $I$ 
based on this circular drawing. 
\qed
\end{proof}
}
\end{lemma}

By Lemma~\ref{I=I'}, we have the following fact 
that tells about the relationship between two problems \MLCMPP CROSSING 
and CIC. 

\begin{corollary}
\label{MLCMPP=CIC}
{\rm
The output of an instance $I$ of \MLCMPP CROSSING is yes 
if and only if the output of an instance $I'$ of CIC. 
}
\end{corollary}

\noindent
{\bf Second Step:} 
Corollary~\ref{MLCMPP=CIC} ensures the correctness of the reduction 
from \MLCMPP CROSSING to CIC. 
However, we notice that the drawing in CIC is restricted in the point 
that edges cannot pass across a circle. 
In the next step to reduce CIC to PLANARITY, 
we force this restriction in the reduction. 
To this end, it suffices to construct a graph, which contains $G'$ of $I'$, 
can be planar only if it contains $G'$ and should be drawn 
within a cycle that corresponds to a circle in a drawing of $G'$ of 
CIC\@. 
To attain this, we adopt $K_4$, which is a minimal non-outerplanar graph.

Now we construct an instance $I''$ of PLANARITY in the following way.
For an instance $I'=(G',\delta')$ of CIC obtained 
from an instance $I$ of \MLCMPP CROSSING, 
we `pad' vertex $1_{\uparrow}$ of $G'$ onto each vertex of $K_4$ 
which is drawn on the plane without crossings, 
and we make it $G''$ of an instance $I''$ of PLANARITY 
(Fig.~\ref{k4_embedding} in Appendix). 

\begin{figure}[htbp]
 \begin{center}
\scalebox{0.65}{\includegraphics{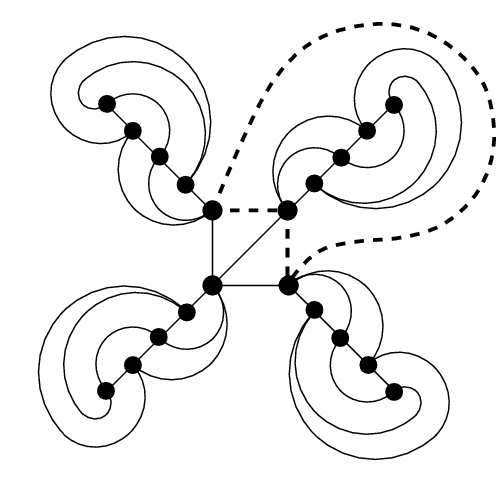}}
 \end{center}
\caption{A graph $G''$ for PLANARITY obtained by a graph $G'$ 
of CIC, where $G'$ is in Fig.~\ref{I->I'}(d). 
A dashed cycle corresponds to the circle in CIC.} 
\label{k4_embedding}
\end{figure}

Then we have the following lemma with respect to an instance $I'$ of 
CIC 
and an instance $I''$ of PLANARITY. 

\begin{lemma}
\label{I'=I''}
{\rm
The output of an instance $I'$ of CIC is yes 
if and only if the output of an instance $I''$ of PLANARITY is yes. 

\begin{proof}
Since $K_4$ is planar but not outerplanar, 
at least one of its vertex is not on the outer boundary of its planar drawing. 
Therefore, if we assume that $G''$ is planar, 
at least one of four `paddings' of $G'$ onto each vertex of $K_4$ 
has to be drawn inside of a cycle of a planar drawing of $K_4$. 
Then if we regard a cycle as a circle, a subgraph of $G''$ composed by $G'$ 
is a drawing of $G'$ inside a circle without crossings. 
Conversely, if we have a non-crossing drawing of $G'$ of $I'$ within a circle, 
we have a non-crossing drawing of $G''$ of $I''$ 
simply by drawing $K_4$ without crossings and padding such a drawing of $G'$ 
onto each vertex of $K_4$ so that their edges do not cross with already 
drawn edges. 
\qed
\end{proof}
}
\end{lemma}

\noindent
Combining Lemmas~\ref{I=I'} and \ref{I'=I''}, 
we see that the output of an instance $I$ of \MLCMPP CROSSING is yes 
if and only if the output of an instance $I''$ of PLANARITY is yes, 
and this shows the correctness of the reduction in two steps. 


Based on the reductions, we show by an algorithm how we can solve 
\MLCMPP CROSSING. 
In lines 1--2 of the algorithm, since it generates $O(|V|)$ vertices 
and $O(|E|+|{\cal L}|)$ edges 
to construct $I'$ from $I$ and then $I''$ from $I'$, 
it requires $O(|V|+|E|+|{\cal L}|)$ time. 
Then for solving PLANARITY for $I''$ in line 3, 
by using a linear-time algorithm \cite{hopcroft+tarjan74}, 
it takes $O(|V|+|E|+|{\cal L}|)$ time. 
Therefore, the overall computational time of this algorithms is 
$O(|V|+|E|+|{\cal L}|)$, and this completes the proof of 
Theorem~\ref{MLCM-PP_crossing_in_P}. 

\begin{algorithm}
\begin{algorithmic}[1]
\renewcommand{\algorithmicrequire}{\textbf{Input:}}
\renewcommand{\algorithmicensure}{\textbf{Output:}}
\renewcommand{\baselinestretch}{0.85}\selectfont
\small
\caption{\bf Algorithm \MLCMPP CROSSING}
\REQUIRE an underlying graph $G=(V,E)$ and a line set ${\cal L}$.
\ENSURE if there exists a layout of ${\cal L}$ on $G$ without crossings YES, 
otherwise NO. 
\STATE construct an instance $I'=(G',\delta')$ of \CIC from $I=(G, {\cal L})$; 
\STATE construct an instance $I''=(G'')$ of PLANARITY from $I'$; 
\STATE solve PLANARITY for $I''$ by using an existing algorithm; 
 \IF{the output for $I''$ is yes}
   \RETURN  yes;
 \ELSE
  \RETURN no;
 \ENDIF
\end{algorithmic}
\end{algorithm}

\section{An Exact Exponential Algorithm for \MLCMPP}
\label{MLCM-P_exact}

In this section, we propose a fast exponential exact algorithm 
for MLCP-P\_PATH which is NP-hard. 

To solve \MLCMPP exactly, a naive approach is to compute the number 
of crossings for all possible layouts of lines in ${\cal L}$ 
by using an $O(|{\cal L}||V|)$ time algorithm for MLCM-PA \cite{bekos08}, 
and then to output the layout with minimum number of crossings among them. 
Since the number of different assignments of two ends (to top or bottom) 
of a line is four, 
this idea yields an algorithm whose running time is 
$O(4^{|{\cal L}|}\times |{\cal L}||V|)$ $= O^*(4^{|{\cal L}|})$.\footnote{%
The $O^*$-notation ignores a polynomial factor, commonly used in the
exponential-time algorithm literature.
}
We propose a faster exponential algorithm 
that works in $O^*(2^{|{\cal L}|})$ time. 

\begin{figure}[htbp]
 \begin{center}
\scalebox{0.63}{\input{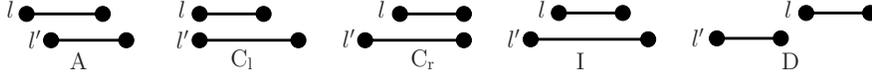}}
 \end{center}
\caption{Five types of relative positions of two lines.} 
\label{2lines_position}
\end{figure}

For the purpose of computing the number of crossings 
when the assignment of lines to top or bottom, we classify the relationship 
of the positions of two lines $l=[i,j]$, $l'=[i',j']$ as follows, that is, 
type A: $i<i'<j<j'$, type ${\rm C_l}$: $i=i'$, type ${\rm C_r}$: $j=j'$, 
type I: $i'<i<j<j'$, and type D: $j<i'$. 
Here we assume without loss of generality that either $j<j'$, 
or $j=j'$ and $i>i'$ holds. 
Fig.~\ref{2lines_position} shows these five types. 
We use Type C to denote both types of ${\rm C_l}$ and ${\rm C_r}$. 

For types A, C and I, 
since two lines of these types have a common interval, 
there are some cases in which crossings of two lines cannot be avoided 
to satisfy the periphery condition, depending on the assignment of line ends. 
On the other hand, MLCM-PA with these assignments of line ends 
can be drawn only with unavoidable crossings \cite{bekos08}, 
the number of those is equal to the minimum number 
of crossings for the given assignment. 

\begin{figure}[htb]
\begin{center}
\scalebox{0.60}{\includegraphics{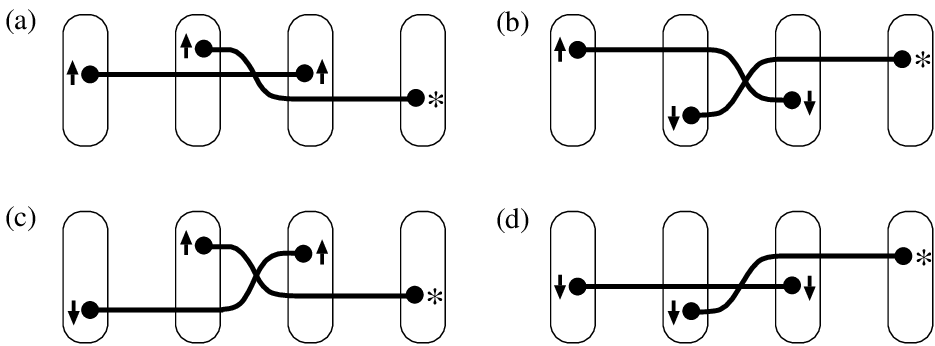}}
\end{center}
\caption{All the assignments of a pair of two lines in type A 
that have a crossing: 
(a) $a_l=(\uparrow, \uparrow)$ and $a_{l'}=(\uparrow, *)$, 
(b) $a_l=(\uparrow, \downarrow)$ and $a_l'=(\downarrow, *)$, 
(c) $a_l=(\downarrow, \uparrow)$ and $a_l'=(\uparrow, *)$, and 
(d) $a_l=(\downarrow, \downarrow)$ and $a_l'=(\downarrow, *)$.}
\label{crossing_assignment}
\end{figure}

Here, we denote an assignment $a_l$ of the left end $i$ and the right end $j$ 
of a line $l=[i,j]$ by a pair of up and down arrows 
$\{\uparrow, \downarrow\}\times\{\uparrow, \downarrow\}$, 
where $\uparrow$ and $\downarrow$ imply top and bottom, respectively. 
For example, if both left and right ends of a line $l$ is assigned to top, 
its assignment is denoted by $a_l=(\uparrow, \uparrow)$. 
We also denote by $a_l=(\uparrow, *)$ to imply 
$a_l=(\uparrow, \uparrow)$ and $a_l=(\uparrow, \downarrow)$. 
The number of different assignments of four line ends of two lines 
is $2^4=16$, and for example for type A (Fig.~\ref{crossing_assignment}), 
we see that the following assignments among those must have a crossing, 
by examining all possible cases. 
We can similarly find these assignments for types C and I, 
and obtain the following observation. 

\begin{fact}
\label{crossing_assignment_ACI}
{\rm
For types A, C and I, a pair of lines 
has a crossing in the following assignments of their ends.

\noindent
Type A: 
(1) $a_l=(\uparrow, \uparrow)$ and $a_{l'}=(\uparrow, *)$, 
(2) $a_l=(\uparrow, \downarrow)$ and $a_l'=(\downarrow, *)$, 
(3) $a_l=(\downarrow, \uparrow)$ and $a_l'=(\uparrow, *)$, or 
(4) $a_l=(\downarrow, \downarrow)$ and $a_l'=(\downarrow, *)$. \\
Type ${\rm C_l}$:
(1) $a_l=(\uparrow, \downarrow)$ and $a_l'=(\downarrow, *)$, or 
(2) $a_l=(\downarrow, \uparrow)$ and $a_l'=(\uparrow, *)$. \\
Type ${\rm C_r}$: 
(1) $a_l=(\uparrow, \downarrow)$ and $a_l'=(*, \uparrow)$, or 
(2) $a_l=(\downarrow, \uparrow)$ and $a_l'=(*, \uparrow)$. \\
Type I:  
(1) $a_l=(\uparrow, \downarrow)$, or 
(2) $a_l=(\downarrow, \uparrow)$. 
}
\end{fact}

The basic idea of the algorithm we propose is 
to try all possible assignments of left ends of lines 
and to greedily determine the assignments of right ends 
for each assignment of left ends 
so that the number of crossings becomes minimum. 
Fact~\ref{crossing_assignment_ACI} guarantees the correctness of 
this greedy strategy. 
We show our algorithm, which we call FixLeftEnd for MLCM-P\_PATH, below. 

\begin{algorithm}
\begin{algorithmic}[1]
\renewcommand{\algorithmicrequire}{\textbf{Input:}}
\renewcommand{\algorithmicensure}{\textbf{Output:}}
\renewcommand{\baselinestretch}{0.82}\selectfont
\small
\caption{\bf Algorithm FixLeftEnd for \MLCMPP}
\REQUIRE an underlying graph $G=(V,E)$ and a line set ${\cal L}$.
\ENSURE assignments of left and right ends of each line 
to $\uparrow$ or $\downarrow$ to minimize the number of crossings. 
\vspace{1mm}
\STATE $main$\{
\STATE $min\_cross=\infty$; 
\STATE $assign(1)$;
\RETURN currently saved assignment; \}
\vspace{1mm}
\STATE $assign$($i$)\{
 \IF{$i=|{\cal L}|+1$}
  \FOR{$j=1$ to $|{\cal L}|$}
   \STATE determine the assignment of right end $t_j$ of line $l_j=[s_j,t_j]$ 
so that the number of crossings with lines $l_k=[s_k,t_k]$ satisfying 
$s_k<t_j<t_k$ becomes minimum (tie breaks by assigning it to top); 
  \ENDFOR
  \STATE crossing\_number = the number of crossings generated 
by the determined assignment; 
  \IF{$crossing\_number < min\_cross$}
   \STATE $min\_cross = crossing\_number;$
   \STATE save the current assignment; 
  \ENDIF
 \ELSE
  \STATE assign the left end of line $l_i$ to top; 
  \STATE $assign({i+1})$;
  \STATE assign the left end of line $l_i$ to bottom; 
  \STATE $assign({i+1})$;
 \ENDIF
\STATE \}
\end{algorithmic}
\end{algorithm}

In line 8 of the algorithm, it determines for each line $l_j$ 
the assignment of the right end of $l_j$ so that it has fewer crossings 
with line $l_k$ (of types A, ${\rm C_l}$ or I) 
whose right end is right to the right end $t_j$ of $l_j$. 
Here, according to Fact~\ref{crossing_assignment_ACI}, 
since whether $l_j$ and $l_k$ cross or not is determined by the assignments 
of both ends of $l_j$ and the left end of $l_k$, 
the assignment of the right end of $l_k$ does not affect it. 
Therefore, to determine the assignment of the right end of $l_j$ in line 8 
does not affect to the crossings with the lines 
whose right ends are left to $t_j$, 
and is not affected by the assignments of right ends of lines 
whose right ends are right to $t_j$. 
This implies that the output by the algorithm has 
minimum number of crossings except the ones caused by pairs of lines 
of type ${\rm C_r}$. 
Notice that the algorithm assigns a right end to top 
when the numbers of crossings caused by its assignment to top and bottom 
are equal. 
Now we can show that any pair of lines of type ${\rm C_r}$ do not cross 
in the output of this algorithm. 

\begin{lemma}
\label{Cr_do_not_cross}
{\rm
Any two lines of type ${\rm C_r}$ do not cross in the output 
of Algorithm FixLeftEnd for MLCM-P\_PATH. 

\begin{proof}
Assume that left ends of two lines $l=[i,j]$ and $l'=[i', j]$ $(i'<i)$ 
of their position are assigned both to top. 
Then according to Fact~\ref{crossing_assignment_ACI}, 
$l$ and $l'$ cross if and only if 
$a_l=(\uparrow,\downarrow)$ and $a_{l'}=(\uparrow,\uparrow)$. 
Let $c_u$ ($c_d$) denote the number of crossings of line $l$ and a line 
in a position of type A, ${\rm C_l}$ or I 
when the right end of $l$ is assigned to top (bottom) by Algorithm FixLeftEnd. 
We define $c_u'$ ($c_d'$) similarly. 
Since the position of $l$ and $l'$ is of type ${\rm C_r}$, 
$c_u \leq c_u'$ and $c_d'\leq c_d$ hold. 
Now if the algorithm assigns right ends of $l$ and $l'$ 
to bottom and top, respectively, so that they cross, 
$c_d< c_u$ and $c_u'< c_d'$ hold. 
This implies $c_u\leq c_u'<c_d'\leq c_d<c_u$, which is a contradiction. 

Similarly to this argument, in cases of left ends of $l$ and $l'$ are 
assigned to top and bottom, bottom and top, and bottom and bottom, 
we can obtain a similar contradiction if the algorithm assigns 
the right ends of $l$ and $l'$ so that they cross. 
Hence, two lines of type ${\rm C_r}$ do not cross 
in the output of the algorithm. 
\qed
\end{proof}
}
\end{lemma}

By Fact~\ref{crossing_assignment_ACI} and Lemma~\ref{Cr_do_not_cross}, 
since determining the assignment of the right end of line $l_j$ greedily 
achieves the minimum number of crossings for a given assignment of left ends, 
we have the following result. 

\begin{theorem}
{\rm
Algorithm FixLeftEnd for MLCM-P\_PATH outputs an assignment of line ends 
that achieves the minimum number of crossings. 
}
\end{theorem}

The assignment of the right end of line $l_j$ is determined 
by counting the numbers of crossings when it is assigned to top or bottom 
and comparing them in $O(|{\cal L}|)\times O(1)$ time. 
Therefore, it takes $O(|{\cal L}|^2)$ time for all the lines. 
Also, since there are $2^{|{\cal L}|}$ ways of assigning 
left ends of all the lines, 
Algorithm FixLeftEnd for MLCM-P\_PATH takes 
$O(2^{|{\cal L}|}\times|{\cal L}|^2)$ $=O^*(2^{|{\cal L}|})$ time in total. 
To obtain an actual layout, 
we can use the algorithm of Bekos et al. \cite{bekos08} for MLCM-PA 
by giving the output of Algorithm FixLeftEnd for MLCM-P\_PATH as its input. 
Thus we have the following theorem. 

\begin{theorem}
{\rm
\MLCMPP is solved in 
$O^*(2^{|{\cal L}|})$ time. 
}
\end{theorem}

\section{Fixed-Parameter Algorithms for MLCM-P\_PATH} 
\label{MLCM-P_fpt}

N\"{o}llenburg \cite{nollenburg10} asked as an open problem if MLCM-P 
is fixed-parameter tractable,
and also pointed out that the `multiplicity' of lines is a possible parameter. 
Here, the {\it multiplicity} of a line set ${\cal L}$ is defined to be 
$\max\{|{\cal L}_{uv}|\ \big|\ (u, v)\in E\}$, 
that is, the maximum number of lines on an edge of an underlying graph. 
In the real metro maps, 
the multiplicity is 
relatively small, therefore fixed-parameter algorithms with respect to 
the multiplicity are considered to run fast. 

Let the multiplicity of an input line set be $k$. 
Then the number of possible permutations of lines on left and right sides 
of each station are both $O(k!)$. 
Since a layout for MLCM-P\_PATH is determined by those permutations 
on both sides of all stations, 
a naive algorithm that enumerates all possible permutations of lines 
on $2|V|-2$ sides of $|V|$ stations 
and outputs a layout with the minimum number of crossings 
after checking if each layout is admissible can run in $O({k!}^{2|V|})$ time. 
However, this is not a fixed-parameter algorithm 
since its running time is not expressed 
in a form of $O(f(k)\cdot |V|^{O(1)})$. 
In this section, we first explain a fixed-parameter algorithm for computing 
an {\it optimal} layout that determines permutations on left and right sides 
of stations from 1 to $n$ by using dynamic programming, 
and then accelerate it by using properties of optimal layouts 
or devising efficient ways of computing.

\subsection{A Naive Fixed-Parameter Algorithm}
\label{fpt_naive}

Let sets of permutations of lines on the left and right sides of station $i$ 
be ${\widehat{\Pi}}_i^\ell$ and ${\widehat{\Pi}}_i^r$, respectively. 
For an input $(G,{\cal L})$ of MLCM-P\_PATH, 
let ${\cal L}_i^r$ $(={\cal L}_{i+1}^\ell)$ denote the set of lines 
that pass edge $(i,i+1)$, 
and denote by ${\cal T}_i^\ell$ and ${\cal T}_i^r$ 
the set of lines that have their right ends on the left side 
and left ends on the right side, respectively, of station $i$. 
We now define two functions $f_\ell$ and $f_r$ 
with respect to the minimum number of crossings of a layout:
\medskip
\begin{description}
  \setlength{\parskip}{-0.8cm} 
  \setlength{\itemsep}{0cm} 
\item[$f_\ell(\pi, i):$] 
the minimum number of line crossings from station 1 to $i$ 
when the lines in ${\cal L}_i^\ell$ are in the order of 
$\pi \in {\widehat{\Pi}}_i^\ell$ on the left side of station $i$; 
  \setlength{\parskip}{-0cm} 
  \setlength{\itemsep}{0cm} 
\item[$f_r(\pi, i):$] 
the minimum number of line crossings from station 1 to $i$ 
when the lines in ${\cal L}_i^r$ are in the order of 
$\pi\! \in\! {\widehat{\Pi}}_i^r$ on the right side of station $i$. 
  \setlength{\parskip}{-1cm} 
\end{description}

\noindent
Then the minimum number of crossings for MLCM-P\_PATH is denoted by 
$\min\{f_\ell(\pi, n)\mid\pi\in {\widehat{\Pi}}_n^\ell\}$. 

We can confirm the substructure optimality property of MLCM-P\_PATH 
as follows.
When the lines in ${\cal L}_i^\ell$ are in the order of 
$\pi \in {\widehat{\Pi}}_i^\ell$ on the left side of station $i$, 
let permutations from on the right side of station 1 to the right side of 
station $i-1$ in ${\widehat{\Pi}}_1^r$, ${\widehat{\Pi}}_2^\ell,\ldots$, 
${\widehat{\Pi}}_{i-1}^\ell$, ${\widehat{\Pi}}_{i-1}^r$ 
that achieves $f_\ell(\pi, i)$ be 
$\pi_1^r$, $\pi_2^\ell,\ldots$, $\pi_{i-1}^\ell$, $\pi_{i-1}^r$, respectively. 
Here $f_\ell(\pi, i)$ is the sum of the number of crossings caused by 
$\pi_1^r$, $\pi_2^\ell,\ldots$, $\pi_{i-1}^\ell$, $\pi_{i-1}^r$ 
to the left side of station $i-1$, 
and by $\pi_{i-1}^r$ on the right side 
of station $i-1$ and $\pi$ on the left side of station $i$. 
If $\pi_1^r$, $\pi_2^\ell,\ldots$, $\pi_{i-1}^\ell$ 
from on the left side of station 1 to the left side of station $i-1$ 
do not achieve $f_r(\pi_{i-1}^r, i-1)$, 
then there exist other permutations that achieve it, 
and replacing $\pi_1^r$, $\pi_2^\ell,\ldots$, $\pi_{i-1}^\ell$ with them 
decreases $f_\ell(\pi, i)$. 
Therefore $\pi_1^r$, $\pi_2^\ell,\ldots$, $\pi_{i-1}^\ell$ 
must be permutations 
that achieve $f_r(\pi_{i-1}^r, i-1)$. 

Similarly, when the lines in ${\cal L}_i^r$ are in the order 
of $\pi \in {\widehat{\Pi}}_i^r$ on the right side of station $i$, 
let permutations from on the right side of station 1 to the left side of 
station $i$ in ${\widehat{\Pi}}_1^r$, ${\widehat{\Pi}}_2^\ell,\ldots$, 
${\widehat{\Pi}}_{i-1}^r$, ${\widehat{\Pi}}_{i}^\ell$ 
that achieve $f_r(\pi, i)$ be 
$\pi_1^r,\pi_2^\ell, \ldots, \pi_{i-1}^r, \pi_i^\ell$, respectively. 
Then, since $f_r(\pi, i)$ $=f_\ell(\pi_i^\ell, i)$ holds 
for a layout to be admissible, 
$\pi_1^r$, $\pi_2^\ell,\ldots$, $\pi_{i-1}^\ell$, $\pi_{i-1}^r$
must be permutations from the right side of station 1 to the right side 
of station $i-1$ that achieve $f_\ell(\pi_i^\ell, i)$. 

As we observed, since MLCM-P\_PATH has a substructure optimality, 
we can derive the following recurrence with respect to $f_\ell$ and $f_r$. 

\vspace{-8pt}
\begin{align}
f_r(\pi, i) &= \begin{cases}
      0 & (\mbox{if } i=1), \\
      \min\{f_\ell(\pi',i)\mid \pi_{{\cal L}_i^r\backslash {\cal T}_i^r} 
      = \pi'_{{\cal L}_i^\ell\backslash {\cal T}_i^\ell} \} & (\mbox{if } i>1),
		\end{cases} \\
f_\ell(\pi, i) &= \min\{f_r(\pi',i-1)+t(\pi', \pi)\}. \label{a}
\end{align}

\noindent
Here, for a permutation $\pi$ of the lines in ${\cal L}$ 
and a subset $A$ $(\subset {\cal L})$, 
we define $\pi_A$ to be a subsequence of $\pi$ 
obtained by deleting the lines of ${\cal L}\setminus A$ from $\pi$. 
In equation (1), ${\cal L}_i^r\backslash {\cal T}_i^r$ denotes 
the set of lines that pass the right end of station $i$ 
except the lines whose left end is on the right side of station $i$, 
and $\pi_{{\cal L}_i^r\backslash {\cal T}_i^r}$ is a permutation 
consists of those lines. 
(Similarly for ${\cal L}_i^\ell\backslash {\cal T}_i^\ell$ 
and $\pi_{{\cal L}_i^\ell\backslash {\cal T}_i^\ell}$.) 
Therefore, $\pi_{{\cal L}_i^r\backslash {\cal T}_i^r} 
= \pi'_{{\cal L}_i^\ell\backslash {\cal T}_i^\ell}$ implies 
that lines do no cross within stations so that layouts are admissible. 
In equation (2), $t(\pi, \pi')$ denotes the number of inversions 
of $\pi$ and $\pi'$, that is, the number of pairs $(i,j)$ 
that satisfy $\pi(i)<\pi(j)$ and $\pi'(i)>\pi'(j)$. 

We now estimate the complexity of computing $f_r$ and $f_\ell$ 
based on these recurrences. 
Let the multiplicity be $k$.
Then $|{\widehat{\Pi}}_i^r|\leq k!$ 
and $|{\widehat{\Pi}}_i^\ell|\leq k!$ hold. 
For computing $f_r$ for a single station, since it checks if the condition 
for $\min$ is satisfied in $O(k)$ time for each of $O(k!)$ $f_\ell$'s, 
it takes $O(k\cdot k!)$ time. 
Similarly, computing $f_\ell$ for a single station 
takes $O(k^2\cdot k!)$ time, 
since it computes the number of inversions in $O(k^2)$ time 
for each of $O(k!)$ $f_r$'s. 
Finally, since there are $O(k!)$ $f_r$'s and $f_\ell$'s for every station 
and there are $|V|$ stations, the overall time becomes 
$O(k^2 (k!)^2|V|)=O(2^{O(k\log k)}|V|)$. 
This dynamic programming approach leads to a fixed-parameter algorithm 
with respect to the multiplicity $k$ 
whose running time is $O(2^{O(k\log k)}|V|)$.

\subsection{Accelerate the Fixed-Parameter Algorithm}

The naive algorithm explained in Subsect.~\ref{fpt_naive}
considered permutations that do not satisfy the periphery condition 
on both sides of stations. 
Still, there may exist other permutations 
that are not necessary for finding an optimal layout. 
In this subsection, we show a series of lemmas that can restrict permutations 
to be considered for deriving an optimal layout. 
Furthermore, we design a faster fixed-parameter algorithm 
by implementing an efficient way of computing a DP table. 



\begin{lemma}
\label{cross_immediate_before}
{\rm
Among optimal layouts for MLCM-PA when underlying graphs are paths, 
there exists one that satisfies the following condition: 
if two lines $l=[i,j]$ and $l'=[i',j']$ ($j<j'$) cross, 
the crossing occurs on edge $(j-1,j)$ of its underlying graph. 
}
\end{lemma}

This is shown by the fact that the algorithm which solves MLCM-PA 
\cite{bekos08} outputs a layout satisfying this property. 
By using Lemma~\ref{cross_immediate_before}, 
we can show a similar property on an optimal layout for MLCM-P\_PATH. 

\begin{lemma}
\label{cross_right_end_MLCMPP}
{\rm
Among optimal layouts for MLCM-P\_PATH,  
there exists one that satisfies the condition that, 
if two lines $l=[i,j]$ and $l'=[i',j']$ ($j$ $<j'$) cross, 
the crossing occurs on edge $(j-1,j)$ of its underlying graph. 

\begin{proof}
An optimal layout for an instance $(G,{\cal L})$ of MLCM-P 
is also that of an instance $(G,{\cal L},{\cal A})$ of MLCM-PA 
where ${\cal A}$ is the corresponding assignment of line ends of that layout. 
By Lemma~\ref{cross_immediate_before}, 
there exists a layout in which crossing two lines 
$l=[i,j]$ and $l'=[i',j']$ ($j<j'$) cross on edge $(j-1,j)$ 
among optimal layouts for $(G,{\cal L},{\cal A})$. 
Since an optimal layout for an instance $(G,{\cal L},{\cal A})$ 
of MLCM-PA is also an optimal layout for $(G,{\cal L})$, the lemma holds. 
\qed
\end{proof}
}
\end{lemma}

The algorithm to solve MLCM-PA when underlying graphs are paths \cite{bekos08} 
determines a layout of each line from its left end to right end. 
This works correctly if we change the direction, that is, 
if it determines from right end to left end. 
This and Lemma~\ref{cross_right_end_MLCMPP} lead to the following corollary. 

\begin{corollary}
\label{cross_left_end_MLCMPP}
{\rm
Among optimal layouts for MLCM-P\_PATH,  
there exists one that satisfies the condition that, 
if two lines $l=[i,j], l'=[i',j']$ ($i$ $<i'$) cross, 
the crossing occurs on edge $(i',i'+1)$ of its underlying graph. 
}
\end{corollary}

Based on Lemma~\ref{cross_right_end_MLCMPP} 
and Corollary~\ref{cross_left_end_MLCMPP}, 
we have the following lemma with respect to crossings in optimal layouts. 

\begin{figure}[hbt]
 \begin{center}
\scalebox{0.65}{\input{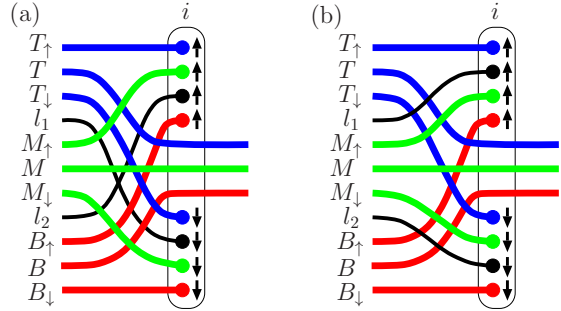}}
 \end{center}
\caption{(a) Relative positions of lines in ${\cal L}_{i-1,i}$, and 
(b) those after changing assignment of right ends of $l_1$ and $l_2$. 
Black lines are $l_1$ and $l_2$, 
blue lines are $T$, $T_\uparrow$ and $T_\downarrow$, 
green lines are $M$, $M_\uparrow$ and $M_\downarrow$, and 
red lines are $B$, $B_\uparrow$ and $B_\downarrow$.} 
\label{typeC_classify}
\end{figure}

\begin{lemma}
\label{C_do_not_cross}
{\rm
In optimal layouts for MLCM-P\_PATH, 
two lines of type C (${\rm C_l}$ and ${\rm C_r}$) do not cross. 

\begin{proof}
First consider type ${\rm C_r}$. 
Assume that two lines $l_1$ and $l_2$ whose right ends are on station $i$ 
cross in an optimal layout. 
Then by Fact~\ref{crossing_assignment_ACI}, 
assignments of their right ends are different. 
Also by Lemma~\ref{cross_right_end_MLCMPP}, 
we may assume that any two lines cross on edge 
immediately before the right end 
of a line whose right end is left to the other's. 

Now we classify the lines in ${\cal L}_{i-1,i}\setminus\{l_1,l_2\}$ as follows 
(Fig.~\ref{typeC_classify}(a)). 
For the lines passing through station $i$, 
let those above $l_1$ and $l_2$ on the right side of station $i-1$ be $T$, 
those between $l_1$ and $l_2$ be $M$, and those below $l_1$ and $l_2$ be $B$. 
For the lines whose assignments of right ends are top, 
let those above $l_1$ and $l_2$ on the right side of station $i-1$ 
be $T_\uparrow$, those between $l_1$ and $l_2$ be $M_\uparrow$, 
and those below $l_1$ and $l_2$ be $B_\uparrow$. 
For the lines whose assignments of right ends are bottom, 
we define $T_\downarrow$, $M_\downarrow$ and $B_\downarrow$ similarly. 

Then we can see, in Fig.~\ref{typeC_classify}(a), 
that the crossings created by $l_1$ and $l_2$ in an optimal layout are 
those by $l_1$ and $l_2$, 
by $l_1$ and lines in $M_\uparrow$, $M$, $B_\uparrow$ or $B$, 
and by $l_2$ and $T$, $T_\downarrow$, $M$ or $M_\downarrow$. 
Therefore, the number of crossings $c_1$ becomes 
$c_1=1+|M_\uparrow|+|M|+|B_\uparrow|+|B|+|T|
+|T_\downarrow|+|M|+|M_\downarrow|$. 
On the other hand, if we change the assignments of right ends of 
both $l_1$ and $l_2$ (Fig.~\ref{typeC_classify}(b)), 
the crossings created by $l_1$ and $l_2$ are 
those by $l_1$ and lines in $T_\downarrow$ or $T$ 
and by $l_2$ and $B_\uparrow$ or $B$, and thus the number of crossings $c_2$ 
becomes $c_2=|T_\downarrow|+|T|+|B_\uparrow|+|B|$. 
Then $c_1-c_2= 2|M| + |M_\uparrow|+|M_\downarrow|+1$, 
and the number of crossings decreases at least by 1, 
which contradicts the assumption that the original layout is optimal. 
Therefore, lines of type ${\rm C_r}$ do not cross in an optimal layout. 

Next consider type ${\rm C_l}$. 
Again assume that $l_1$ and $l_2$ whose left ends are on station $i$ 
cross in an optimal layout. 
Then by Fact~\ref{crossing_assignment_ACI}, 
assignments of their left ends are different. 
Also by Lemma~\ref{cross_right_end_MLCMPP}, 
we may assume that any two lines cross on edge immediately after the left end 
of a line whose left end is right to the other's. 
Now by flipping lines in ${\cal L}_{i,i+1}$ horizontally, 
we can do similar arguments as above, 
and finally lines of type ${\rm C_l}$ do not cross in an optimal layout. 
\qed
\end{proof}
}
\end{lemma}

\begin{figure}[htbp]
 \begin{minipage}{0.38\hsize}
  \begin{center}
\scalebox{0.63}{\includegraphics{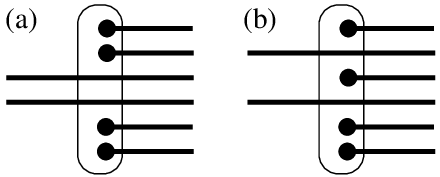}}
 \end{center}
 \caption{A permutation $\pi$ on the right side of a station: 
(a) $P_r(\pi)=1$, and (b) $P_r(\pi)=0$.} 
\label{Pr}
 \end{minipage}
\begin{minipage}{0.02\hsize}
\mbox{}
\end{minipage}
 \begin{minipage}{0.58\hsize}
  \begin{center}
\scalebox{0.63}{\includegraphics{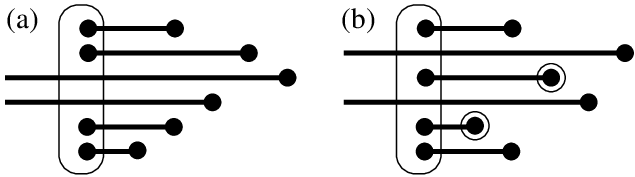}}
 \end{center}
 \caption{A permutation $\pi$ on the right side of a station: 
(a) $Q_r(\pi)=1$, and (b) $Q_r(\pi)=0$. 
Circled line ends are between the other lines.} 
\label{Qr}
 \end{minipage}
\end{figure}

We introduce some notation about the properties 
satisfied by permutations of lines. 
For a permutation $\pi$ of lines in ${\cal L}_i^r$ on the right side 
of station $i$, we define a function $P_r(\pi)$ as follows: 
$P_r(\pi)=1$ if left ends of all lines of ${\cal T}_i^r$ 
$(\subseteq {\cal L}_i^r)$ satisfy the periphery condition; 
$P_r(\pi)=0$, otherwise (Fig.~\ref{Pr}). 
We define similarly a function $P_\ell(\pi)$ for a permutation $\pi$ 
of lines of ${\cal L}_i^\ell$ on the left side of station $i$. 
For a permutation $\pi$ of lines in subset $A$ $(\subset{\cal L})$, 
we define a function $Q_r(\pi)$ as follows: 
$Q_r(\pi)=1$ if all right ends of lines of $A$ satisfy 
the periphery condition when they are layout in parallel 
in the order of $\pi$; 
$Q_r(\pi)=0$, otherwise (Fig.~\ref{Qr}). 
We define similarly a function $Q_\ell(\pi)$ for a permutation $\pi$ 
of lines in $A$, that is, 
$Q_\ell(\pi)=1$ if all left ends of lines in $A$ satisfy 
the periphery condition when they are layout in parallel 
in the order of $\pi$; $Q_\ell(\pi)=0$, otherwise. 

By Lemmas~\ref{cross_right_end_MLCMPP} and \ref{C_do_not_cross}, 
we restrict the permutations on the left and right side of stations 
to be considered in order to obtain optimal layouts. 
We first restrict permutations on the left side of each station $i$. 
Due to the periphery condition, a permutation $\pi$ on the left side 
of a station satisfies $P_\ell(\pi)=1$. 
By Lemma \ref{C_do_not_cross}, since lines in ${\cal T}_i^\ell$ do not cross 
in optimal layouts, only permutations satisfying 
$Q_\ell(\pi_{{\cal T}_i^\ell})=1$ need to be considered. 
In addition, by Lemma~\ref{cross_right_end_MLCMPP}, 
since we may assume that lines passing through station $i$ do not cross 
before station $i$, we have only to consider permutations that satisfy 
$Q_\ell(\pi_{{\cal L}_i^\ell\setminus {\cal T}_i^\ell})=1$ 
to find an optimal layout. 
We let the set of permutations that satisfy these conditions 
be ${\Pi_i^\ell}$, that is, ${\Pi_i^\ell}=\{\pi\mid 
P_\ell(\pi)=1, Q_\ell(\pi_{{\cal T}_i^\ell})=1,
Q_\ell(\pi_{{\cal L}_i^\ell\setminus {\cal T}_i^\ell})=1\}$. 

We then restrict permutations on the right side of each station $i$. 
Due to the periphery condition, a permutation $\pi$ on the right side 
satisfies $P_r(\pi)=1$. 
By Lemma \ref{C_do_not_cross}, since lines in ${\cal T}_i^r$ do not cross 
in optimal layouts, only permutations satisfying 
$Q_r(\pi_{{\cal T}_i^r})=1$ need to be considered. 
In addition, by Lemma~\ref{cross_right_end_MLCMPP}, 
since we can assume that lines passing through station $i$ do not cross 
before station $i$, we have only to consider permutations that satisfy 
$Q_\ell(\pi_{{\cal L}_i^r\backslash {\cal T}_i^r})=1$
to find an optimal layout. 
We let the set of permutations that satisfy these conditions be ${\Pi_i^r}$, 
that is, ${\Pi_i^r}=\{\pi\mid P_r(\pi)=1, Q_r(\pi_{{\cal T}_i^r})=1,
Q_\ell(\pi_{{\cal L}_i^r\backslash {\cal T}_i^r})=1\}$. 

Now we can change the recurrence equation (2) in Subsect.~\ref{fpt_naive} 
in the following form by incorporating these restrictions on permutations 
on the left and right sides of every station:

\vspace{-8pt}
\begin{equation}
f_\ell(\pi, i) = \min\{f_r(\pi',i-1)+t(\pi', \pi)\mid
\pi_{{\cal L}_i^\ell\backslash {\cal T}_i^\ell} 
= \pi'_{{\cal L}_{i-1}^r\backslash T_i^\ell}, 
\pi_{{\cal T}_i^\ell} = \pi'_{{\cal T}_i^\ell} \}. 
\end{equation}

\noindent
In equation (3), 
$\pi_{{\cal L}_i^\ell\backslash {\cal T}_i^\ell} 
= \pi'_{{\cal L}_{i-1}^r\backslash T_i^\ell}$ implies 
by Lemma~\ref{cross_right_end_MLCMPP} that lines 
that pass through station $i$ do not cross, 
and $\pi_{{\cal T}_i^\ell} = \pi'_{{\cal T}_i^\ell}$ implies 
by Lemma~\ref{C_do_not_cross} that lines that have their right ends on 
the left side of station $i$ do not cross. 
Now we have the following lemma with respect to the sizes of 
$\Pi^\ell_i$ and $\Pi^r_i$. 

\begin{lemma}
\label{permutation_set_size}
{\rm
Let $k$ be the multiplicity of lines. 
Then $|{\Pi_i^\ell}|\leq 2^k$ and $|{\Pi_i^r}|\leq 2^k$ $(1\le i\le n)$ hold. 

\begin{proof}
First we consider any permutation $\pi\in {\Pi_i^\ell}$. 
Since $Q_\ell(\pi_{{\cal L}_i^\ell\backslash {\cal T}_i^\ell})=1$, 
lines in ${\cal L}_i^\ell\backslash {\cal T}_i^\ell$ 
are sorted by their positions of left ends 
and are located in this order at the left side of station $i$ 
either above or below line $l$ whose left end is the leftmost among them. 
Such ways correspond to those of partitioning lines in 
${\cal L}_i^\ell\backslash {\cal T}_i^\ell$ (except $l$) 
into two subsets, and the number is not greater 
than $2^{|{\cal L}_i^\ell\backslash {\cal T}_i^\ell|}$. 
Also, since $P_\ell(\pi)=1$ and $Q_\ell(\pi_{{\cal T}_i^\ell})=1$, 
lines in ${\cal T}_i^\ell$ are sorted by their positions of left ends 
and are located in this order at the left side of station $i$ 
either above of below lines in ${\cal L}_i^\ell\backslash {\cal T}_i^\ell$. 
Such ways correspond to those partitioning lines in ${\cal T}_i^\ell$ 
into two subsets, and the number is not greater than $2^{|{\cal T}_i^\ell|}$. 
Thus the size of ${\Pi_i^\ell}$ is not greater than 
$2^{|{\cal L}_i^\ell\backslash {\cal T}_i^\ell|}
\times 2^{|{\cal T}_i^\ell|} = 2^{|{\cal L}_i^\ell|}$. 
Since ${|\cal L}_i^\ell|\leq k$, we have $2^{|{\cal L}_i^\ell|}\leq 2^k$. 

Next we consider any permutation $\pi\in {\Pi_i^r}$. 
Since $Q_\ell(\pi_{{\cal L}_i^r\backslash {\cal T}_i^r})=1$, 
lines in ${\cal L}_i^r\backslash {\cal T}_i^r$ 
are sorted by their positions of left ends 
and are located in this order at the right side of station $i$ 
either above or below line $l$ whose left end is the leftmost among them. 
Such ways correspond to those of partitioning lines in 
${\cal L}_i^r\backslash {\cal T}_i^r$ (except $l$) into two subsets, 
and the number is not greater 
than $2^{|{\cal L}_i^r\backslash {\cal T}_i^r|}$. 
Also, since $P_r(\pi)=1$ and $Q_r(\pi_{{\cal T}_i^r})=1$, 
lines in ${\cal T}_i^r$ are sorted by their positions of right ends 
and are located in this order at the right side of station $i$ 
either above of below lines in ${\cal L}_i^r\backslash {\cal T}_i^r$. 
Such ways correspond to those partitioning lines in ${\cal T}_i^r$ 
into two subsets, and the number is not greater than $2^{|{\cal T}_i^r|}$. 
Thus the size of ${\Pi_i^r}$ is not greater than 
$2^{|{\cal L}_i^r\backslash {\cal T}_i^r|}\times 2^{|{\cal T}_i^r|}$ 
$= 2^{|{\cal L}_i^r|}\leq 2^k$. 
\qed
\end{proof}
}
\end{lemma}

\noindent
According to Lemma~\ref{permutation_set_size}, 
we can estimate, by a similar argument to the one in Subsect.~\ref{fpt_naive}, 
that $f_r$ and $f_\ell$ can be computed by recurrence equations (1) and (3) 
in $O(k^2 4^k|V|)$ time. 

We further improve this time complexity by computing the recurrence forward, 
that is, by generating only possible permutations from left to right, 
starting from $\pi \in {\Pi_1^r}$ to compute $f_r(\pi, 1)$. 
We consider possible permutations of lines in ${\cal L}_{i+1}^\ell$ 
on the left side of station $i+1$ that can be generated from $\pi$, 
where $\pi$ is a permutation of lines in ${\cal L}_i^r$ 
on the right side of station $i$. 
By Lemma~\ref{cross_right_end_MLCMPP}, 
they are exactly the permutations where lines in ${\cal T}_{i+1}^\ell$ 
are deleted from $\pi$ and lines in ${\cal T}_{i+1}^\ell$ are added 
either to the head or the tail of $\pi$. 
By Lemma~\ref{C_do_not_cross}, since lines of type C do not cross 
in optimal layouts, the number of ways to add lines in ${\cal T}_{i+1}^\ell$ 
either to the head or the tail is $|{\cal T}_{i+1}^\ell|+1\leq k+1$ 
(determine the position for division of ${\cal T}_{i+1}^\ell$ 
by keeping its order in $\pi$, as shown in Fig.~\ref{generate_left}). 

Next, we consider possible permutations of lines in ${\cal L}_i^r$ 
on the right side of station $i$ that can be generated from $\pi$, 
where $\pi$ is a permutation of lines in ${\cal L}_i^\ell$ 
on the left side of station $i$. 
Since ${\cal L}_i^r=({\cal L}_i^\ell\backslash {\cal T}_i^\ell)
\cup {\cal T}_i^r$ and lines do not cross inside of a station 
(to be admissible), 
they are exactly the permutations where lines in ${\cal T}_{i}^\ell$ 
are deleted from $\pi$ and lines in ${\cal T}_{i}^r$ are added 
either to the head or the tail of $\pi$. 
Again, by Lemma~\ref{C_do_not_cross}, since lines of type C do not cross 
in optimal layouts, the number of ways to add lines in ${\cal T}_{i}^r$ 
either to the head or the tail of $\pi$ is $2^{|{\cal T}_i^r|}$ 
(partition ${\cal T}_{i+1}^\ell$ into two, and add one to the head 
and the other to the tail with avoiding crossings of lines of type C, 
as shown in Fig.~\ref{generate_right}). 

Here we define a binary relation $R$ on the set of permutations 
on the left side of station $i$ 
by $(\pi,\pi')\in R \Longleftrightarrow$ 
${\pi}_{{\cal L}_i^\ell\backslash {\cal T}_i^\ell}$ 
$={\pi'}_{{\cal L}_i^\ell\backslash {\cal T}_i^\ell}$ 
($\pi, \pi'\in\Pi_i^\ell$). 
Then $R$ is an equivalence relation, 
and all permutations on the right side of station $i$ 
generated from permutations (on the left side of station $i$) 
in an equivalent class $[\pi]$ whose representative is $\pi\in\Pi_i^\ell$ 
are the same. 
Among permutations on the right side of station $i$, 
a permutation that achieves minimum number of crossings can be generated 
from a permutation $\pi^*$ that satisfies 
$f_\ell(\pi^*, i)=\min\{f_\ell(\pi, i)\mid \pi\in[\pi^*]\}$ 
in each equivalent class. 
Therefore, we do not need to generate permutations for all permutations 
on the left side of station $i$ but for the ones that make $f_\ell$ minimum 
in each equivalent class. 
Notice, by Lemma~\ref{permutation_set_size}, that the number of different 
equivalent classes is $2^{|{\cal L}_i^\ell\backslash{\cal T}_i^\ell|}$. 

\begin{figure}[htbp]
 \begin{minipage}{0.48\hsize}
  \begin{center}
\scalebox{0.63}{\input{./generate_left.tps}}
  \end{center}
  \caption{Generate permutations on the left side of station $i+1$ 
from ones on the right side of station $i$.} 
  \label{generate_left}
 \end{minipage}
\begin{minipage}{0.02\hsize}
\mbox{}
\end{minipage}
 \begin{minipage}{0.48\hsize}
  \begin{center}
\scalebox{0.63}{\input{./generate_right.tps}}
  \end{center}
  \caption{Generate permutations on the right side of station $i$ 
from ones on the left side of station $i$.} 
  \label{generate_right}
 \end{minipage}
\end{figure}

We then estimate the complexity for computing each $f_\ell$ and $f_r$ 
in the algorithm. 
First, we consider the time for computing $f_\ell$ 
by generating possible permutations on the left side of station $i+1$ 
from the ones on the right side of station $i$. 
Since each permutation $\pi\in\Pi_i^r$ generates $O(k)$ permutations 
in $O(k)$ time for each and computes the number of inversions with $\pi$ 
in $O(k^2)$ time, 
all possible permutations on the left side of station $i+1$ for each $\pi$ 
can be generated in $O(k \times (k+k^2))$ $=O(k^3)$ time. 
Since $|\Pi_i^r|\leq 2^k$ by Lemma~\ref{permutation_set_size}, 
computing $f_\ell$ for station $i+1$ by generating permutations 
from all permutations on the right side of station $i$ 
takes $O(2^k\times k^3)=O(k^32^k)$ time. 

Next, we consider the time for computing $f_r$ 
by generating possible permutations on the right side of station $i$ 
from the ones on the left side of station $i$. 
When we compute $2^{|{\cal L}_i^\ell\backslash{\cal T}_i^\ell|}$ 
equivalent classes and a permutation that minimizes $f_r$ in each class, 
we assume that all classes are stored in the lexicographic order of lines, 
and therefore we have access to each class 
in $O(\log2^{|{\cal L}_i^\ell\backslash{\cal T}_i^\ell|})$ 
$=O(|{\cal L}_i^\ell\backslash{\cal T}_i^\ell|)=O( k)$ time 
by using a binary search. 
Hence computing equivalent classes takes $O(k2^k)$ time. 
We then generate $2^{|{\cal T}_i^r|}$ possible permutations on the right side 
of station $i$ in $O(k)$ time for each of 
$2^{|{\cal L}_i^\ell\backslash{\cal T}_i^\ell|}$ equivalent classes. 
Therefore, it takes 
$O(k2^k$ $+2^{|{\cal L}_i^\ell\backslash{\cal T}_i^\ell|}$ 
$\times k2^{|{\cal T}_i^r|})$ $=O(k2^k)$ time 
for generating all possible permutations on the right side of station $i$ 
from permutations on the left side of station $i$. 
Finally, since there are $|V|$ stations, the algorithm takes overall 
$O((k^32^k + k2^k)|V|) = O(k^3 2^k|V|)$ time 
for computing all $f_\ell$'s and $f_r$'s, 
and we have the following theorem. 
\begin{theorem}
\label{fpt}
{\rm
\MLCMPP is fixed-parameter tractable 
with respect to the multiplicity $k$ of lines. 
It is solved in $O(k^3 2^k|V|)$ time. 
}
\end{theorem}

\section{Conclusion}
\label{conclusion}

In this paper, we studied an NP-hard problem MLCM-P\_PATH, 
that is, MLCM-P whose underlying graph of the input is a path.
On this problem, we show that its decision version 
belongs to P, 
and proposed an $O^*(2^{|{\cal L}|})$-time exact algorithm. 
We also proposed a fixed-parameter algorithm that runs 
in $O(k^3 2^k|V|)$ time with respect to the multiplicity $k$ of lines. 
This implies that \MLCMPP is fixed-parameter tractable, 
and partially solves the open problem posed in \cite{nollenburg10} 
affirmatively. 

It is still an important open problem
if MLCM-P is fixed-parameter tractable for more general input graphs, 
such as planar graphs. 
Furthermore, MLCM-P may be fixed-parameter tractable 
with respect to the other parameters than the multiplicity of lines. 
For example, for the problem to ask if there is a layout 
with line crossings no more than $k$, 
the fact that the decision version belongs to P 
may help to show 
that it is fixed-parameter tractable
with respect to the minimum number of crossings $k$. 
We also point out that it is significant 
to determine if MLCM-P CROSSING for general graphs belongs to P\@.

\end{document}